# Gain and Efficiency Enhancement in Free Electron Laser by means of Modulated Electron Beam

**Vivek Beniwal[1], Suresh C. Sharma[2] and S. Hamaguchi[3]**

[1]Department of Physics, Maharaja Surajmal Institute of Technology, C-4, Janakpuri, New Delhi 110 0058, **India**

[2]Department of Physics, GPMCE (G.G.S.Indraprastha University, Delhi), **India**

[3] Science and Technology Center for Atoms, Molecules and Ions Control, Graduate School of Engineering, Osaka University, 2-1 Yamada-oka, Suita, Osaka 565-0871, **Japan**

## ABSTRACT

Beam premodulation on free electron laser (FEL) offers considerable enhancement in gain and efficiency when the phase of the premodulated beam is $-\pi/2$ and the beam is highly modulated implying the maximum beam oscillatory velocity due to wiggler. The growth rate of the FEL instability increases with the modulation index and reaches maximum when the modulation index $\Delta \sim 1\cdot 0$.

**Keywords :** FEL, Wiggler, Gain and Efficiency

## 1. INTRODUCTION

The free electron laser (FEL) is a fascinating device for an efficient generation of high power coherent radiation in a wide band of frequencies[1-3]. Recently, there has been growing interest in studying the free electron laser (FEL)[4-5] by prebunched electron beams.

## 2. INSTABILITY ANALYSIS

Consider the interaction of a FEL with a static magnetic wiggler,

$$\vec{B}_w = \hat{x} B_w e^{ik_w z}, \qquad (1)$$

where $k_w = -|k_w|$ is the wiggler wave vector.

A premodulated relativistic electron beam of density $n_{ob}$, velocity $v_{ob}\hat{z} = \left\{2\frac{eV_b}{m}(1+\Delta\sin\omega_o\tau)\right\}^{1/2}$, relativistic gamma factor $\gamma_0 = 1+\frac{eV_b}{m_0c^2}(1+\Delta\sin\omega_0\tau)$ propagates through the interaction region (cf. Fig. 1),

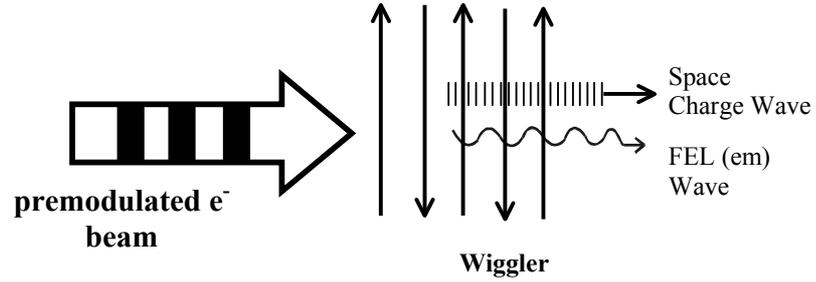

**premodulated e⁻ beam**

**Wiggler**

Space Charge Wave

FEL (em) Wave

**Fig. 1 Schematic of a Free Electron Laser**

where $\Delta\ (=V_1/V_b)$ is the modulation index, $V_b$ is the beam voltage, $m_0c^2$ is the rest mass energy of the electrons, e and m are the electron charge and mass, $\omega_o(\simeq k_{oz}v_{ob})$ and $k_{0z}$ are the modulation frequency and wave number of the premodulated electron beam, respectively. Moreover, $\omega_o\tau = \psi = \omega_o\left(t-\frac{z}{v_{ob}}\right)$ is the phase of the premodulated beam. It couples a beam space charge mode $(\omega, \vec{k})$ and an electromagnetic wave $(\omega_1, \vec{k}_1)$,

$$\vec{E}_1 = \hat{y}\ E_1\ e^{-i(\omega_1 t - k_1 z)}, \qquad (2)$$

$$\vec{B}_1 = c\ \frac{\vec{k}_1 \times \vec{E}_1}{\omega_1},$$

$$\phi = \phi\ e^{-i(\omega t - kz)},$$

where $\omega_1 = \omega$, $k_1 = k + |k_w|$. The coupling may be viewed as a parametric process involving a wiggler (o, $k_w$), a negative beam mode $(\omega, \vec{k})$: $\omega = k\ v_{ob} - \omega_{pb}/\gamma_0^{2/3}$, $\omega_{Pb} = (4\pi\ n_{ob}\ e^2/m)^{1/2}$ is the beam plasma frequency, and a radiation wave $\omega_1 \simeq k_1 c$. The phase matching conditions yield, the frequency of the electromagnetic radiation

$$\omega_1 = 2\gamma_0^2 \left( |k_w| v_{ob} - \frac{\omega_{pb}}{\gamma_0^{3/2}} \right). \quad (3)$$

The beam electrons acquire the transverse wiggler velocity is obtained as

$$\vec{v}_w = -\hat{y} \frac{eB_w}{imck_w \gamma_0 (1 + \Delta \sin \omega_0 \tau)}. \quad (4)$$

Under the influence of an electromagnetic perturbing mode $(\omega_1, \vec{k}_1)$, the electrons acquire a transverse velocity.

$$\vec{v}_1 = \frac{e\vec{E}_1}{im\omega_1 \gamma_0 (1 + \Delta \sin \omega_0 \tau)}. \quad (5)$$

The wiggler and the radiation wave exerts a ponderomotive force on the beam electrons at $(\omega, \vec{k})$

$$\vec{F}_p = -\frac{e}{2c}(\vec{v}_1 \times \vec{B}_w + \vec{v}_w \times \vec{B}_1) = ie\vec{k}\phi_p. \quad (6)$$

where

$$\Phi_p = \frac{-eE_1 B_w}{2cm\omega_1 \gamma_0 (1 + \Delta \sin \omega_0 \tau)k_w}. \quad (7)$$

and

$$v_w = \frac{eB_w}{mc\gamma_0 (1 + \Delta \sin \omega_0 \tau)k_w} \simeq v_{osc}. \quad (8)$$

Under the influence of the ponderomotive force and the self consistent field $E = -\nabla \phi$, the electrons acquire an axial velocity

$$v_{pz} = -\frac{ek(\phi + \phi_p)}{m\gamma_0^3 (\omega_1 - kv_{ob})}. \quad (9)$$

The resulting density perturbation $n_p$ can be obtained by solving the equation of continuity and is given as

$$n_p = -\frac{n_{ob} ek^2 (\phi + \phi_p)}{m\gamma_0^3 (\omega_1 - kv_{ob})^2}. \quad (10)$$

Using the density perturbation $n_p$ in the poisson's equation, we obtain

$$\epsilon \phi = -\chi_e \phi_p \quad , \tag{11}$$

where $\epsilon = 1 + \chi_e = 1 - \dfrac{\omega_{pb}^2}{\gamma_0^3 (\omega_1 - kv_{ob})^2}$ , and $\chi_e$ is the beam susceptibility.

The second harmonic nonlinear current density at $(\omega_2, \vec{k}_2)$ is given by

$$\vec{J}_2 = -n_{ob} e \vec{v}_2 - \frac{1}{2} n_p e \vec{v}_1, \tag{12}$$

Using $\vec{J}_2$ in the wave equation

$$k_2^2 \vec{E}_2 - \frac{\omega_2^2}{c^2} \vec{E}_2 = 4\pi \frac{i\omega_2}{c^2} \vec{J}_2 \quad , \tag{13}$$

we obtain the nonlinear dispersion relation

$$(\omega_1^2 - k_1^2 c^2)(\omega_1 - kv_{ob})^2 = \frac{\omega_{Pb}^2 e k^2 v_w E_1 \left(1 - \dfrac{\chi_b}{\epsilon}\right)}{8\gamma_0^4 m \omega_1 (1 + \Delta \sin \omega_0 \tau)}. \tag{14}$$

## RAMAN REGIME

At high beam currents ($\omega_{pb} > \Gamma$, where $\Gamma$ is the growth rate ) one may achieve $\epsilon \approx 0$, i.e., $\omega \simeq k\, v_{ob} - \omega_{pb}/\gamma_0^{3/2}$. In this limit self consistent potential far exceeds the ponderomotive one ($\Phi >> \Phi_p$). Equation (14) takes the form

$$(\omega_1^2 - k_1^2 c^2)\left[(\omega_1 - kv_{ob})^2 - \frac{\omega_{pb}^2}{\gamma_0^3}\right] = \frac{\omega_{pb}^2 e k^2 v_w E_1}{8\gamma_0^4 m \omega_1 (1 + \Delta \sin \omega_0 \tau)} \quad . \tag{15}$$

One looks for a solution of Eq. (15) around the simultaneous zeros of the left hand side. The first factor when equated to zero, gives the radiation mode while the second one gives the beam space charge mode. We write

$$\omega_1 = k_1 c + \delta = k_{0z} v_{ob} - \frac{\omega_{pb}}{\gamma_0^{3/2}} + \delta$$

and by solving Eq. (15) for $\delta$, the growth rate turns out to be

$$\Gamma = \frac{1}{4}\left[\frac{\omega_{pb}e^2k^2B_wE_1}{2m^2\omega_1^2\gamma_0^{7/2}ck_w(1+\Delta\sin\omega_0\tau)^2}\right]^{1/2}, \quad (16)$$

$$|\delta_r| = +\Gamma/\sqrt{3}.$$

## GAIN ESTIMATE

Following Liu and Tripathi[3], the gain 'G' can be determined by expanding P and $\psi$ to different orders in A.

$$\langle\Delta P\rangle = \langle-(P_1+P_2)\rangle_{\xi=1} = A^2 G$$

$$= \frac{A^2}{P_o^2}\left[1-\cos P_o - \frac{P_o}{2}\sin P_o\right] = \frac{-A^2}{8}\frac{d}{dx}\left[\frac{\sin^2 x}{x^2}\right] \quad (17)$$

where $x = \frac{P_o}{2} = \frac{(\gamma_o - \gamma_r)\omega_1 L}{(\gamma_r^2-1)^{3/2} 2c}$, $A = \frac{L^2\omega_1 eE_p}{2mc^2(\gamma_r^2-1)^{3/2}}$.

and $-\frac{d}{dx}\left[\frac{\sin^2 x}{x^2}\right]$ is a gain function.

## EFFICIENCY

Following Liu and Tripathi[3], the efficiency in the Raman regime is

$$\eta \sim \frac{\gamma_0^3}{(\gamma_0-1)}\frac{\Gamma}{\omega_1} = \frac{\gamma_0^3}{4\omega_1(\gamma_0-1)}\left[\frac{\omega_{pb}e^2k^2B_wE_1}{2m^2\omega_1^2\gamma_0^{7/2}ck_w(1+\Delta\sin\omega_0\tau)^2}\right]^{1/2}. \quad (18)$$

## 3. RESULTS

In the calculations we have used parameters for the experiment of Cohen *et al*[4-5] and the corresponding parameters are : electron beam energy $E_b$ =0.07 MeV, beam current $I_b$=1.2A, and beam cross-section $A_b$ = 0.126 cm$^2$, wiggler field $B_w$ = 300 Gauss, wiggler wavelength $\lambda_w$ =4.42 cm, the modulation frequency of the premodulated beam $\omega_0$ =4.25GHz, electric field $E_1$ = 50 esu. In Fig. 2, we have plotted the variation of the growth rate (in rad. sec$^{-1}$) in the Raman regime [using Eq. (16)] with modulation index $\Delta$ when the phase of the premodulated electron beam is $-\pi/2$, i.e., when the beam is in the retarding zone. The growth rate of the FEL instability increases with the modulation index and has the largest value when $\Delta \sim 1{\cdot}0$ . For $\Delta=0$, the value of the growth rate $\Gamma \sim 1{\cdot}64 \times 10^8$ rad. sec$^{-1}$. In Fig. 3, we have plotted the Gain versus frequency (in GHz). The parameters are same as Fig. 2 and for the interaction length L = 100 cm. We can see from the Fig. 3 , that our theoretical plot is similar to the experimental observation of Cohen *et al*[4] . In our theoretical calculation the gain is found around 0.18 at frequency 4·5 GHz. In the experimental observation of Cohen *et al*[4], the gain increases with frequency and reaches maximum $\sim 0{\cdot}18$ at frequency $\sim 4{\cdot}539$ GHz. Hence our theoretical results qualitatively and quantitatively are similar to the experimental observation of Cohen *et al*[4]. As the modulation index increases, the growth rate of the FEL instability increases and this implies the enhancement in the efficiency of FEL devices [cf. Eq. (18)].

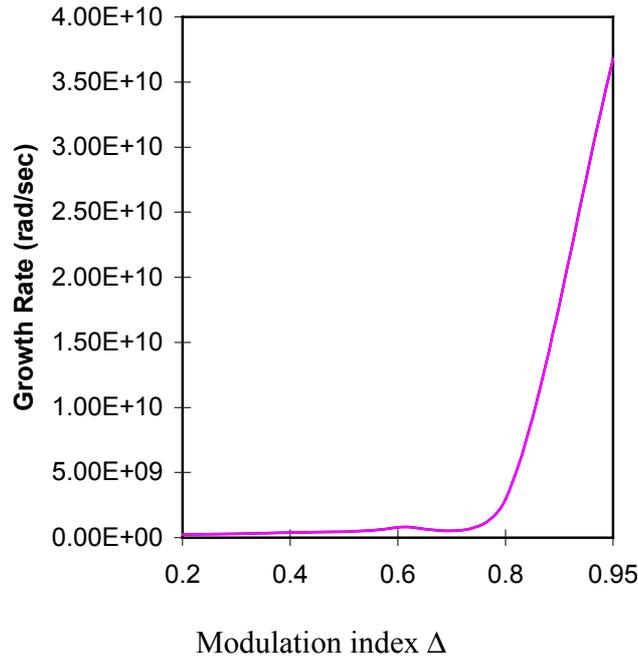

Fig. 2  Growth rate $\Gamma$ (in rad sec$^{-1}$) as a function of modulation Index $\Delta$

for the parameters are given in the text and when $\sin \omega_0 \tau = -1$.

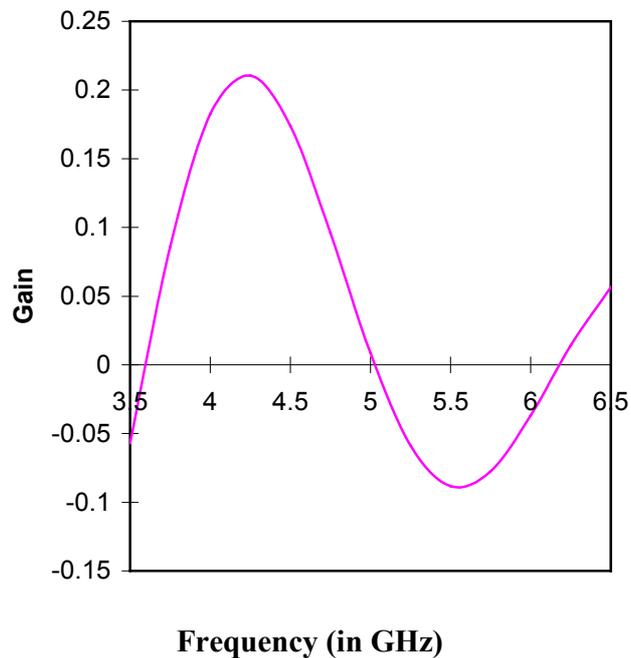

**Frequency (in GHz)**

Fig. 3  Gain versus frequency (in GHz) for the same parameters as Fig. 2

and for the length of the interaction region L = 100 cm.